\documentclass[12pt,preprint]{aastex}

\shorttitle{solar chemical composition}
\shortauthors{Lin et al.}

\begin{document}

\title{Seismic study of the chemical composition of the solar convection zone}

\author{Chia-Hsien Lin}
\affil{Astronomy Department, Yale University,
       P.O. Box 802101, New Haven, CT 06520-8101, U.S.A.}

\author{H.~M.~Antia}
\affil{Tata Institute of Fundamental Research, Homi Bhabha Road,
       Mumbai 400005, India}

\and

\author{Sarbani Basu}
\affil{Astronomy Department, Yale University,
       P.O. Box 802101, New Haven, CT 06520-8101, U.S.A.}

\begin{abstract}
Recent downward revision of solar heavy-element abundances using
three-dimensional atmospheric model has introduced  serious
discrepancies between standard solar models and helioseismic inferences
about  solar structure.
In this paper, we investigate the possibility of determining
the heavy-element abundances
using helioseismic inversion techniques with
the hope of providing an independent estimate.
We use the adiabatic index,
$\Gamma_1 \equiv (\partial \ln P/\partial \ln \rho)_s$,
as a probe
to examine the effects of the total heavy-element abundance,
as well as the effects due to the abundance of individual elements.
Our inversion results show that
the new, lower, abundance increases the discrepancy between the Sun and
the solar models.

\end{abstract}


\section{Introduction}

Until recently, 
various analyses had shown that
the current standard solar models are in good agreement with
the real solar structure to the order of $10^{-3}$
\citep[see, e.g.,][]{1995RvMP67.781B,Gough1996Sci272.1296G,2001ApJ555.990B,2002RvMP74.1073C,2003ApJ599.1434C,2003ApJ583.1004B,2004SoPh220.243R}.
These standard solar models are usually constructed with
OPAL equation of state \citep{1996ApJ456.902R,2002ApJ576.1064R},
the OPAL opacity \citep{1996ApJ464.943I},
and the heavy-element abundances from Grevesse \& Sauval (1998) or \citet{GN1993}.
Recently, 
by using a three-dimensional hydrodynamical model atmosphere
and improved atomic and molecular data,
\citet{2002ApJ573L.137A,AGSAPK2004, AGS2005} revised
the photospheric abundances for the most abundant heavy elements.
The new abundances are
 markedly lower than the earlier values of
\citet{GS1998} or \citet{GN1993}, and result in 
 a total metallicity,  $Z=0.0122$,
which is about 30\% lower than the GS98 value  of $Z=0.0169$.
The standard solar models implemented with the revised abundances deviate
significantly from the solar structure as determined by 
the helioseismic analysis.
In particular,
the convection zone is too shallow and
the helium abundance in the convection zone too low,
which lead to poor match of the sound-speed and density profiles in the Sun
\citep[e.g.,][]{BA2004ApJ606L.85B,BBPS2005ApJ618.1049B,guzik2005ApJ627.1049G,2004PhRvL.93u1102T}.
These models also have a different core-structure than the Sun (Basu et al.~2007).
The discrepancy has
triggered a great deal of  interest in re-examining solar models.
Various groups have explored the possibility of
adjusting the opacity, diffusion rate, and neon abundance of the models
to see if the discrepancy caused by the low  abundances might be resolved.
However, the attempts have not been successful
\citep[e.g.,][]{BA2004ApJ606L.85B,BBPS2005ApJ618.1049B,guzik2006,2004PhRvL.93u1102T,2006ApJ.644.1292A,basu-2007}.
Even invoking a different composition in the radiative interior through
accretion \citet{astroph0611619} fail to remove the discrepancy.

An independent spectroscopic analysis by \citet{2006ApJS..165..618A},
which uses a thermal profile different from
the one in \citet{AGSAPK2004},
suggests an oxygen abundance that is closer to the previous value 
(i.e., Grevesse \& Sauval 1998; henceforth GS98) than the most recent value  (Asplund, Grevesse
\& Sauval 2005; henceforth AGS05).
On the other hand, using 3-dimensional models with free parameters,
Socas-Navarro \& Norton (2007) find an oxygen abundance which is slightly
lower than \citet{AGSAPK2004}. Thus there appears to be considerable
uncertainty in spectroscopic determinations of abundances, and it would
be interesting to attempt an independent determination of abundances
using the seismic data.
\citet{2006ApJ.644.1292A} carried out a helioseismic analysis
to determine the heavy-element abundance.
Their results suggest that $Z=0.0172\pm0.002$,
which is in good agreement with the value of \citet{GS1998}.
In that study, the abundance was estimated
by using a dimensionless
sound-speed gradient $W(r) = (r^2/Gm)(d c^2/dr)$,
which deviates from the ideal-gas value of $-2/3$ in the ionization zones
of elements.
The deviation can then be used to measure the heavy-element abundances.
However, because the ionization zones of different heavy elements overlap,
it is difficult to estimate their abundances separately. Hence, this analysis
was used to estimate only the total heavy-element abundance, $Z$.

In this paper,
we conduct a systematic study of
the effects of both the total abundance and the element mixture
by using another dimensionless quantity,
the adiabatic index $\Gamma_1 \equiv (\partial \ln P/\partial \ln \rho)_s$
($s$ being the entropy).
Our aim is to first identify
the signatures due to the variation of the element abundance
and then use this knowledge to assess the abundance difference between
the Sun and the models. 

Since the normally used equations of state (EOS),
such as OPAL \citep{2002ApJ576.1064R}
and MHD
\citep{DAM1987ApJ319.195D,DMHM1988ApJ332.261D,HM1988ApJ331.794H,MDH1988ApJ331.815M},
do not allow the user to change the chemical element mixture,
and are currently available only with a fixed mixture,
they are not suitable for our investigation.
Instead, following  \citet{2006ApJ.644.1292A},
we use the \citet{EFF1973} equation of state with Coulomb
corrections  (Guenther et al. 1992; Christensen-Dalsgaard \& D\"appen 1992).
This equation of state (referred to as CEFF, the `Coulomb-corrected' EFF) includes all
 ionization states of 20 elements.
The ionization fractions are calculated using only the ground state partition function.
With the CEFF EOS we can specify the relative heavy-element
abundance that we want to use, and hence, study the influence of different
heavy-element mixtures.
Studies have also shown that it gives 
a reasonable description of the thermodynamic structure of the solar material.

The reason for choosing the adiabatic index $\Gamma_1$ 
to probe the chemical composition is that it
is sensitive to both the total abundance and the element mixture.
Although $\Gamma_1$ also depends on temperature and density, these dependence
can be separated from compositional variations if the EOS is known.
The disadvantage of using $\Gamma_1$ is that the errors 
in the EOS also affect $\Gamma_1$.
The effects of the EOS errors and the composition errors
might be entangled and difficult to separate.
To systematically study the relation among 
$\Gamma_1$, EOS and the chemical composition,
we constructed 
a number of test models with different abundances and EOS.
The test models and solar data used in this study are described in \S~\ref{sec:data},
the inversion method is explained in \S~\ref{sec:inversion}, and
the results from the theoretical study and from using solar data are
presented in \S~\ref{sec:results}. Finally, \S~\ref{sec:summary}
summarizes the results of this study.

\section{Solar data and models} \label{sec:data}

We use solar oscillation frequencies obtained from data collected
by the Michelson Doppler Imager (MDI) on board SOHO during its
first 360 days of observations -- 1996 May 1 to 1997 April 25.
We refer to this as the MDI360 set
\citep{schouetal1998}.
We have chosen this set, since among the available helioseismic data
for intermediate degree modes,
this  set was analyzed from the longest time series. 
Other sets
have been obtained from much shorter, 72 or 108 days, time series.
Solar oscillation frequencies are known to change with activity, and
a longer time series would involve  observations from periods of 
increasing solar activity, which would change the frequencies.
It is a well-established that solar oscillations frequencies increase with solar activity. 
However, it is also known that the increase is such that it does not reflect a 
change in structure of the solar interior (Basu 2002).
The MDI360 set was obtained from a
time series when solar activity was low. This helps minimize solar-cycle related
frequency shifts, while providing lower errors than the 72 day sets.
From the estimated errors in our results it is clear that the errors
are acceptable for our work. It may be possible to improve the results 
by individually inverting all 72 day sets available for the
entire solar cycle   and then averaging the
results. 

To study the effects of the chemical composition,
we use a series of models constructed with 
different compositions.
Specifically,
we constructed
several models that have the same relative mixture of heavy-element 
abundances (either GS98 or AGS05) but different $Z/X$,
ranging from 0.0165 to 0.0245, to examine the effect of metallicity.
We also made models 
that have the same $Z/X$ but with the abundance of some  chemical elements either increased
or lowered.
It should be noted that the opacity tables or EOS for a solar model are
calculated with a specified mixture of heavy elements that does
not change with depth. The value of $Z$ on the
other hand
 is determined by specified value of $Z/X$ at the surface and varies
with depth due to diffusion. To increase the relative abundance of one
element with respect to the rest without changing $Z/X$,  the
abundances of other elements is reduced to account for this adjustment
in the mixture. Another alternative is to increase the abundance of
one element and increase $Z/X$ appropriately to keep the abundance of other
elements at standard values. In some cases (e.g., for Neon) we have tried
both options and the results can be compared.
These models are static models calculated by solving the
equations of stellar structure at a fixed age. Inputs needed are
the hydrogen and heavy element abundance profiles (though not the absolute
abundances). We have used the   hydrogen and heavy element
composition profiles from a solar model of Brun et al.~(2002). The
model incorporates  diffusion of helium and heavy
elements and mixing in the tachocline region.
The composition profiles are scaled to the required values of $X$ and $Z$.
The exact structure of the models is not crucial for our work since
we are looking at the intrinsic $\Gamma_1$ difference that depends only on
the EOS and heavy element abundances. This difference does not
depend on the difference in structure between the Sun and the
model, nor does it depend on the difference in the helium abundance.
Consequently, we do not make any special effort
to get models with correct sound speed profile or the convection zone
depth or the helium abundances. Nevertheless, these quantities for all
models used in this work are listed in Table~1. In this table $r_b$ is the
position of the base of the convection zone. These models use the
OPAL opacity table computed for the appropriate heavy element mixture
and use low temperature opacities from Kurucz (1991).
The purpose of these models is to examine the signatures in the 
adiabatic index of differences in the relative mixture of heavy elements.
Since $\Gamma_1$ reflects the effects of both the equation of state and
the chemical composition,
we constructed two models implemented with the latest version of the OPAL equation of state.
These models are referred to as 245OPAL and 165OPAL, and 
have $Z/X=0.0245$ and $0.0165$ respectively. Because they
are constructed with the OPAL EOS, the heavy-element mixture
contains only four elements (C, N, O, Ne). In order 
to examine the effects resulting from the difference between CEFF and OPAL EOS,
we have constructed  two CEFF models that use the same mixture and $Z/X$ as the OPAL models.
These models with OPAL mixtures use the standard OPAL opacity tables to
ensure that the difference is only due to EOS and not due to opacities.
The properties of all models used in this paper 
are summarized in Table~\ref{tab:ceffmdl}.

The relative difference in $\Gamma_1$ can be determined  by helioseismic inversion.
Although the complete set of oscillation modes can be computed for these models,
we only used the modes that are also present in the observed data
for the inversions between models in order to get realistic results.
For the purpose of assessing the effect of the errors in the observed frequencies,
we added random errors that are consistent with those in the observed frequencies
to the model frequencies.
The observational data set used for our mode selection is
the aforementioned MDI360 data set.

\section{Inversion procedures} \label{sec:inversion}
The frequencies of solar oscillation modes depend on
the solar structure.
The starting point of helioseismic  inversions  is the
linearization of the oscillation equations around a known solar model (the so-called
reference model) using the variational principle.
The frequency differences can then be related to the relative variations
in sound speed ($c$) and density ($\rho$) between either two models or between the
Sun and the reference model.
The  relation between the differences in frequency and 
these two variables (i.e., $c$ and $\rho$)
can be written as \citep{DPS1990MNRAS244.542D,AB1994A&AS107.421A}:
\begin{eqnarray}
\frac{\delta \omega_i}{\omega_i} &=&
\int_0^R K^i_{c^2, \rho}(r) \frac{\delta c^2}{c^2}(r) dr +
\int_0^R K^i_{\rho, c^2}(r)
     \frac{\delta \rho}{\rho}(r) dr +
\frac{F_{\rm surf}(\omega_i)}{Q_i} + \epsilon_i,
\label{eqn:inv}
\end{eqnarray}
where
$c$ is the adiabatic sound speed,
$K^i$ are the  kernels, 
and $\epsilon_i$ is the observational error in $\delta\omega_i/\omega_i$.
$F_{\rm surf}(\omega_i)/Q_i$
represents the effect
of uncertainties in the model close to the surface,
and is usually called the ``surface term''.
Here, $Q_i$ is a measure of the mode inertia.

The structure variables, $c^2$ and $\rho$, in Eq.~\ref{eqn:inv} can be
converted into any other pair of independent variables.
Specifically, for our study,
we convert the variables $c^2$ and $\rho$ to
$\Gamma_1$ and $\rho$ by using the relation $c^2=\Gamma_1 P/\rho$ and
the equation of hydrostatic equilibrium,
which links the pressure and density.
While the variable differences in Eq.~\ref{eqn:inv} are the differences calculated 
at the same fractional {\em depth},
we are interested in the difference of $\Gamma_1$ compared at the same $P$ and $\rho$
(or any pair of thermodynamic variables).
It is because comparing $\Gamma_1$ at the same $P$ and $\rho$
removes the effects due to the discrepancies in the main macro-physical properties 
(e.g., the solar structure variables $P$ and $\rho$), and, hence,
such $\delta \Gamma_1/\Gamma_1$ reflects only 
the microphysical (i.e., EOS) and composition discrepancies.
To calculate $\delta \Gamma_1/\Gamma_1$ that represents
only the microphysical discrepancies,
which we call the ``intrinsic'' $\Gamma_1$ difference,
we followed the technique proposed by \citet{BCD1997}:
\begin{eqnarray}
\frac{\delta \Gamma_1}{\Gamma_1} &=&
\left( \frac{\partial \ln \Gamma_1}{\partial \ln P} \right)_{Y,\rho}
\frac{\delta P}{P} +
\left( \frac{\partial \ln \Gamma_1}{\partial \ln \rho} \right)_{Y,P}
\frac{\delta \rho}{\rho} +
\left( \frac{\partial \ln \Gamma_1}{\partial Y} \right)_{P,\rho} \delta Y
+
\frac{\delta \Gamma_{1,{\rm int}}}{\Gamma_1},
\label{eqn:dg1int}
\end{eqnarray}
where
$Y$ is the Helium abundance by mass, and
$\delta \Gamma_{1,\rm int}/\Gamma_1$
is the relative difference caused by
the microphysical discrepancies.
Consequently, the linear relation between $\delta \omega/\omega$ and
$\delta \Gamma_{1,\rm int}/\Gamma_1$ becomes:
\small
\begin{eqnarray}
\frac{\delta \omega_i}{\omega_i} =
\int_0^R K^i_{u, Y}(r) \frac{\delta u}{u}(r) dr +
\int_0^R K^i_{Y, u}(r) \delta Y(r) dr +
\int_0^R K^i_{\Gamma_1, \rho}(r)
     \frac{\delta \Gamma_{1, {\rm int}}}{\Gamma_1}(r) dr +
\frac{F_{\rm surf}(\omega_i)}{Q_i} + \epsilon_i,
\label{eqn:invdg1}
\end{eqnarray}
\normalsize
where
$u\equiv P/\rho$ is the isothermal sound speed. The functions $K^i(r)$ are known
functions of the reference model.

We use the method of Optimally Localized Averages (OLA) to determine
${\delta \Gamma_{1, {\rm int}}/{\Gamma_1}}$ by `inverting'  Eq.~\ref{eqn:invdg1}.
The OLA technique (Backus \& Gilbert 1968) attempts to produce a linear
combination of data such that the resulting resolution kernel or `averaging kernel' is suitably
localized while simultaneously controlling the error estimates. 
Rabello-Soares et al.~(1999) discuss the formulation  of OLA inversion to
study solar structures and  how one may go about determining the
inversion parameters for this type of an inversion.

The ${\delta \Gamma_{1, {\rm int}}/{\Gamma_1}}$ in the formulation above has two contributions,
one from differences in the equation of state, and another from differences
in $Z$:
\begin{equation}
\frac{\delta \Gamma_{1, {\rm int}}}{\Gamma_1}=
\frac{\delta \Gamma_{1, {\rm EOS}}}{\Gamma_1}+  \frac{\partial \ln \Gamma_1}{\partial Z}\delta Z.
\label{eqn:delz}
\end{equation}
In principle, it should be possible to separate out the two parts, however, in practice
we find that the limited number of mode frequencies make it difficult to get a 
reliable separation of the two quantities by inversions. Hence we limit ourselves
to ${\delta \Gamma_{1, {\rm int}}/{\Gamma_1}}$ in this work rather than its two
components.

\section{Results} \label{sec:results}
\subsection{What the models tell us}\label{subsec:models}

The first part of our study involves pairs of models and allows us
to investigate  the individual as well as combined effects
on $\Gamma_{1, \rm int}$ caused by 
 differences in $Z/X$,  the 
relative abundances of different elements, and  the  EOS. This
also allows us to test our inversion technique.
In order to do so, 
we use the frequency differences between
two known models to infer $\delta\Gamma_{1,\rm int}/\Gamma_1$ between them
and then compare it with the known value.
In this way, 
we can systematically investigate features resulting from
different factors (e.g., differences in EOS, metallicity,
relative composition, etc.) and whether these features  can be determined
reliably through inversions. This exercise  enables us to determine whether
we can discern subtle features at any level of significance with the
currently available data sets.
The  knowledge gained from these tests helps to
interpret the features obtained by inverting the frequency differences
between the Sun and a known model. 

\begin{figure}
\centering
\includegraphics[height=0.7\linewidth,angle=90]{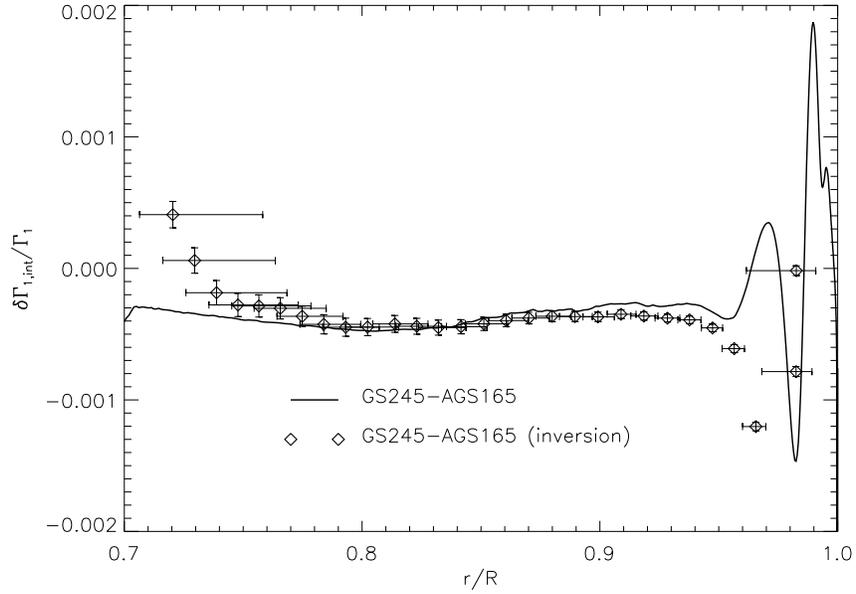}
\caption{
A comparison of the inverted and the exact $\Gamma_{1,\rm int}$ differences
between two models. The line is the exact results, the points with the
error bars are the results obtained by inverting the frequency differences
between the two models using only the modes in  the MDI360 set. The
vertical error bars are a measure of the uncertainties caused by errors in the 
frequencies. The horizontal error bars are a measure of the resolution of
the inversion.
}
\label{fig:invl400}
\end{figure}

In Fig.~\ref{fig:invl400} we compare the exact and inverted $\delta \Gamma_{1,\rm int}/\Gamma_1$
between two models. The inversions were carried out using only those modes
that are present in the   MDI360 mode set. We used the errors on the
observed frequencies to determine the uncertainties in the inversion
results.
We see that we are able to obtain the differences
successfully by inverting the frequency differences between the models. The
inversion results however, are most accurate between about $0.77$ and $0.92 R_\odot$
for the mode set used. The limited number of high-$l$ modes prevents us from 
inverting closer to the surface. This is unfortunate, since, 
the largest differences in $\delta \Gamma_{1,\rm int}/\Gamma_1$ caused by
differences in $Z/X$, EOS, or relative abundances of heavy elements occurs in the
region $r > 0.95R_\odot$. However, we should still be able to make some inferences.
The region $r>0.95R_\odot$ would, in any case, be difficult to use  for this work since
the variation in $\Gamma_1$ in this region are dominated by HeII ionization,
which overwhelms the much smaller signal due to ionization of heavy elements.
The reason we cannot invert deeper than $0.77R_\odot$ is that we need to suppress the
very large signature of $u$ differences before the signature of the $\delta \Gamma_{1,\rm int}/\Gamma_1$ 
differences can be extracted. The difference $\delta u/u$ is an order
of magnitude or more larger just below the base of the convection zone and some
part of the signal leaks into the lower part of the convection zone,
thus making it difficult to suppress $\delta u/u$ contribution reliably in this region.
This problem is more severe if the convection zone depth in the two models
differ significantly.
This results in either large errors, or very poor resolution,
or both. In the rest of the paper, we shall therefore, concentrate only on the 
results for $r > 0.77R_\odot$.

\begin{figure}
\centering
\includegraphics[width=0.8\linewidth]{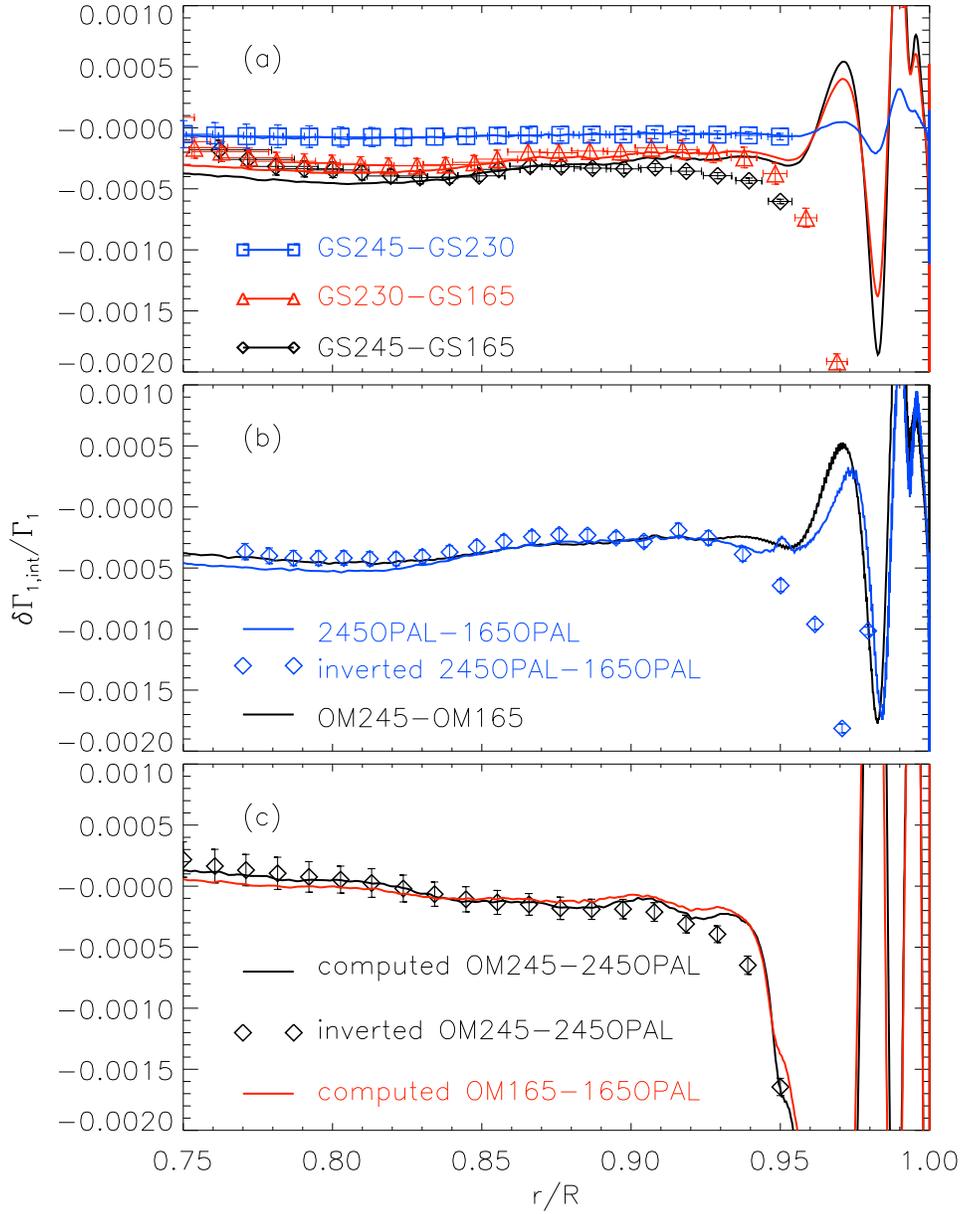}
\caption{ The $\Gamma_{1, \rm int}$ differences between different pairs of
models. Panel (a) illustrates the effect of changing $Z/X$. Panel (b) shows
the response of models with different equations of state to changes
in $Z/X$. Panel (c) shows what happens when the models have the same
$Z/X$ and relative heavy element abundances,  but different equations of state. In all panels the lines
are the exact results, and the points are the results obtained by inverting 
the frequency differences between the pairs of models.
}
\label{fig:f2}
\end{figure}

%%%% Effect of Z/X
In Fig.~\ref{fig:f2}(a) we show $\delta \Gamma_{1,\rm int}/\Gamma_1$ for
models that have the same EOS and relative abundances, but
different values of $Z/X$. We find that the difference in $Z/X$ causes
an almost featureless  parallel shift in the deeper regions ($r < 0.92R_\odot$),
while there are deep depressions and small humps closer to the surface.
The figure also shows the results obtained by inverting the frequency
differences between the models. Unfortunately, the lack of high-$l$
p-modes means that we cannot probe the region that shows the
largest differences caused by $Z/X$. We can only probe the
smaller parallel shifts at $r < 0.92R_\odot$.
We find that for small differences in $Z/X$, as in the
case of the models  GS245 and GS230,
$\delta \Gamma_{1,\rm int}/\Gamma_1$  between the
models below $r\approx 0.92R_\odot$ is comparable to the estimated errors.
In other words, the inversion cannot distinguish between two models if 
the difference in $Z/X$ is less than $\approx 6\%$.

%%%%%%%%Effect of EOS differences
The EOS used to construct our solar models are not perfect and
are known to have differences with respect to the  actual EOS of the solar matter (see
e.g., Basu \& Christensen-Dalsgaard 1997; Basu et al.~1999). Hence, to interpret 
results obtained with solar data, we need to know what type of differences
are seen in $\Gamma_{1, \rm int}$  because of EOS differences alone.
For this purpose, we looked at the difference in $\Gamma_{1, \rm int}$  between
models constructed with the OPAL EOS and CEFF EOS. To ensure that the revealed features 
are solely due to the differences in EOS, we use CEFF models OM245 and OM165,
that have the same heavy-element mixture as
that implemented in OPAL EOS tables. 
Fig.~\ref{fig:f2}(b) shows that for the same difference in $Z/X$ between a pair of models,
the difference in  $\Gamma_{1, \rm int}$ does not depend too much on the
EOS used to construct the pair. The difference between OPAL and CEFF
model pairs is of the order of the inversion errors.
Fig.~\ref{fig:f2}(c) shows the difference in $\Gamma_{1,\rm int}$ between
different EOS for models with the same composition.
We can see that although the difference between the two EOSs 
causes very large features near the surface ($ r>0.93R_\odot$),
the two EOSs are almost indistinguishable below $r\approx0.90R_\odot$.
We can very easily invert the  $\Gamma_{1, \rm int}$ 
differences between the two models in the region that is permissible by our mode sets. The
fact that the two EOSs are indistinguishable throughout most of the convection
zone also means that any solar result obtained using these  CEFF models will
be close to those obtained using OPAL models as long as we do
not study regions very close to the surface.

%%% Results of individual differences 

\begin{figure}
\centering
\includegraphics[height=0.8\linewidth]{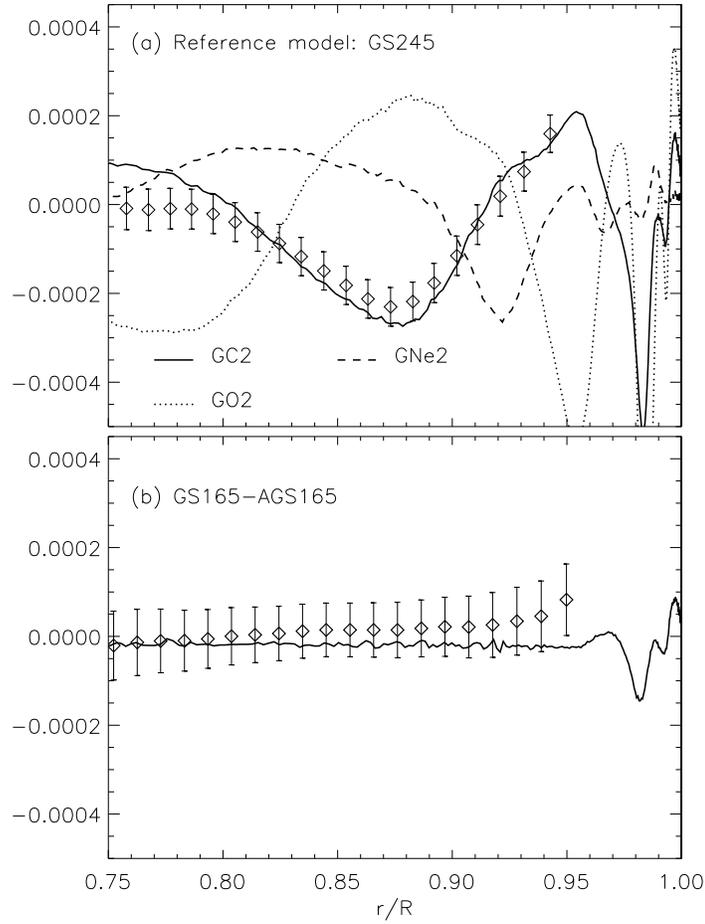}
\caption{
The effect of changes in the relative heavy-element abundances
keeping $Z/X$ the same.
Panel (a) shows the effect when one of the elements is enhanced
by a factor of 2. 
Panel (b) shows the $\Gamma_{1, \rm int}$
differences between two models, both with $Z/X=0.0165$, but one constructed with
the GS98 relative heavy-element abundance and the other the AGS05 relative
heavy-element abundance.
}
\label{fig:f3}
\end{figure}

Fig.~\ref{fig:f3}(a) shows what happens when the abundance of any
one element is increased. As can be seen, the different elements leave a distinct pattern.
The changes shown in the figure are large enough to be seen in the inverted results.
It should be noted that while for the model shown the abundances were changed
by a factor of two, the oxygen abundance in the AGS05 and GS98 tables
differ only by a factor of  about 1.5 and  the carbon abundance differs by a
factor of about 1.35.
 The curve for neon is of interest
since Antia \& Basu (2005)  had suggested that increasing the
neon abundance by a factor of about 4 may bring the solar models constructed with the 
AGS05 abundances to have the correct position of the convection-zone base.
Bahcall et al.~(2005b) suggested a change by factor of between 2.5--3.5 to bring the
convection-zone base and the convection zone helium abundance back into
agreement with solar values. Similar results have been obtained by Zaatri et
al.~(2007) using the scaled small frequency separation in low degree
modes.
We see that adding neon makes the $\delta\Gamma_{1, \rm int}$
positive below $0.9 R_\odot$ and negative between 0.9 and $0.95R_\odot$. 
The results shown in Fig.~\ref{fig:f3}(a) are however, somewhat
disappointing in that it suggests that we could wipe out
detectable changes in $\Gamma_{1, \rm int}$ by, e.g., increasing the abundance of oxygen
and carbon simultaneously. We could also increase the signal by reducing one element
and increasing another,  
in other words, there may be multiple chemical compositions (with same $Z/X$)
that may produce almost
indistinguishable $\Gamma_1$ in the convection zone.
This indicates that 
the chemical composition of the solar convection zone may not be uniquely
determined from $\Gamma_{1, \rm int}$. 
The cancellation effects may be seen
in  the $\Gamma_{1, \rm int}$ differences between AGS05 and GS98 models with the same
$Z/X$ (Fig.\ref{fig:f3}b). Another reason that the AGS05 and GS98 models
with the same $Z/X$ have such similar  $\Gamma_{1, \rm int}$ is that the
relative abundances
of the important elements like C,N,O and Ne between the GS98 and AGS05 abundance
tables is almost the same, since these elements are reduced by almost identical
factors in the AGS05 tables relative to the GS98 values.
In general, the EOS is determined by the number density of different ions and
the relative abundance by number of heavy elements with respect to hydrogen
is of order of $10^{-4}$. Hence the typical differences in $\Gamma_1$ due to
variations in $Z$ are of this order as can be seen from Figs.~1,2. In the
case of difference between GS98 and AGS05 mixtures with the same $Z$, the
differences in relative abundances are much smaller and hence the effect is
also very small.

\subsection{Results with solar data}

\begin{figure}[ht]
\centering
\includegraphics[height=0.8\linewidth,angle=90]{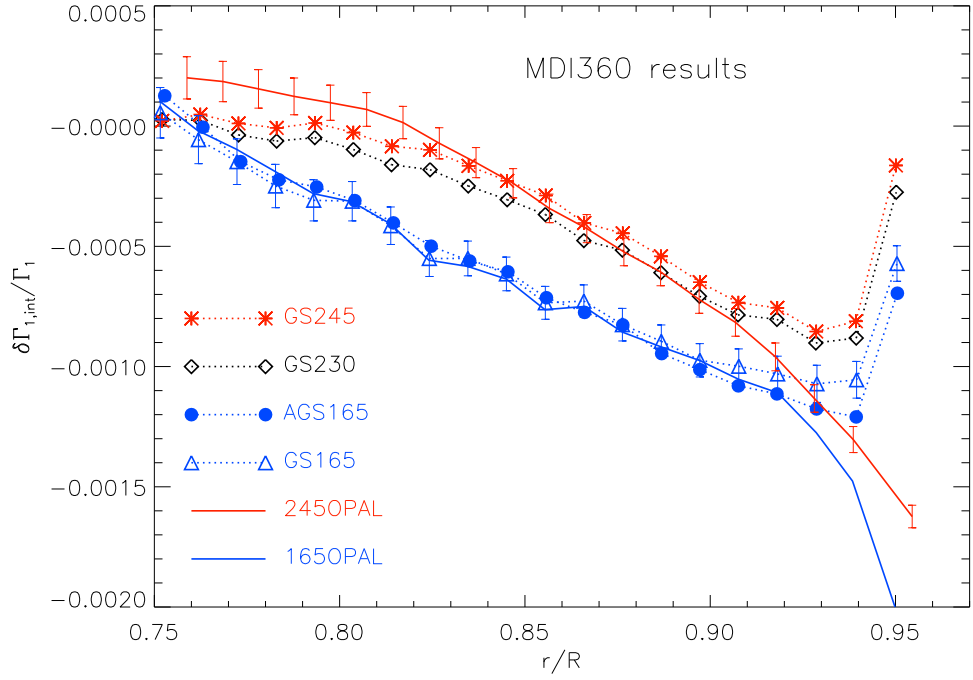}
\caption{Differences in $\Gamma_{1, \rm int}$ between the Sun and different
models obtained by inverting the frequency differences between the Sun and the models.
The models are described in Table~1.
}
\label{fig:f4}
\end{figure}

The inversion results of $\delta \Gamma_{1,\rm int}/\Gamma_1$
between the Sun and various reference models are presented in
the Figures 4, 5 and 6.
The inversion results using the solar data MDI360 and various models
are shown in Fig.~\ref{fig:f4}. The first notable inference
is that models with $Z/X=0.0165$ are more discrepant with respect to the Sun
than models with $Z/X=0.0245$ or $Z/X=0.023$. This is yet another result
that shows that the Sun has a higher $Z/X$ than that given by AGS05.
The difference between the Sun and models with $Z/X=0.0165$ appear to
be the same regardless of whether the models were constructed with the
GS98 or AGS05 relative heavy-element abundances. This is not surprising
given the results  in Fig.\ref{fig:f3}(b) discussed earlier.
Thus using our inversions we cannot say whether the GS98 or AGS05 relative
heavy-element mixture is closer to that in the Sun.
We can only say that the AGS05
model with $Z/X=0.0165$ does not match the solar $\Gamma_1$.
For OPAL models too the agreement with the
Sun is worsened for $Z/X=0.0165$.
It may be noted that in this case the OPAL models use a very different
heavy-element mixture as compared to other models, which would also
account for some of the differences.

Even though the $\delta \Gamma_{1,\rm int}/\Gamma_1$ differences between
the Sun and 
the  models with $Z/X=0.0245$ and $Z/X=0.023$  are visibly different,
the difference in the region $0.77R_\odot <r< 0.90R_\odot$
is comparable to the magnitude of the inversion errors. 
Despite the difference between CEFF and OPAL,
Fig.~\ref{fig:f4} shows that
the solar models implemented with either OPAL or CEFF equation of state favor
the previous heavy-element abundance
(i.e., $Z/X=0.0245$ and $0.0230$) over the latest, lower value, $Z/X=0.0165$.
Hence, based on the precision of current helioseismic data,
models  with either $Z/X=0.0245$ or $Z/X=0.0230$
are equally plausible.

We find that for the same value of $Z/X=0.0245$, enhancing the  carbon abundance
of a model brings it in closer agreement with the Sun, as is shown in
Fig.~5. As can also be seen, enhancing the oxygen abundance actually
worsens the agreement, particularly when compared to a model with  a normal GS98 
relative heavy-element abundance. 
Given that one of the suggestions to improve models constructed with
AGS05 abundances was to increase the neon abundance (Antia \& Basu 2005;
Bahcall et al. 2005b), we turn our attention to models with enhanced neon abundance.
Antia \& Basu (2005) had suggested a neon enhancement by a factor of about 4.
Keeping all other elemental abundances at the AGS05 level, this would
imply a $Z/X$ of $0.0209$. We first look at models with $Z/X=0.0209$ with and
without neon enhancement to disentangle the effects of neon enhancement
from those of increased $Z/X$. The results are shown in Fig.~6(a). What we see 
is that enhancing neon actually worsens the agreement with the Sun in the deeper
part of the convection zone, but makes the agreement better above about $0.90R_\odot$.
However, the neon enhanced models are better than  models with the unmodified AGS05 relative
abundances and the AGS05 total metallicity of $Z/X=0.0165$ as can be
seen in Fig.~6(b). We can also see that the model with neon enhanced
by a factor of two and C,N,O increased by an amount corresponding to their
1$\sigma$ errors in the AGS05 tables (which was another suggestion
of Antia \& Basu 2005) also give a better agreement than the
plain AGS05 model. However, neither of these models fare as well as the
model with GS98 abundances and abundance ratios. Thus it appears that the improvement relative
to AGS05 models
is more due to the increased $Z/X$ in the neon enhanced models, than due to the 
enhancement of neon. Thus while enhancing neon brings the position of the convection-zone 
 base of the solar models in agreement with that of the Sun, the discrepancy in
the adiabatic index remains to a large extent.

\section{Summary}
\label{sec:summary}

In this work we examine the difference in adiabatic index, $\Gamma_1$ between
the Sun and different solar models.  Our aim is to determine whether
this can be used to determine the heavy-element abundances in the convection
zone.
In order to examine the effect of heavy-element abundances on $\Gamma_1$, we use the
so-called `intrinsic' $\Gamma_1$ differences, 
between the Sun and the
models, i.e., differences in $\Gamma_1$ at the same pressure, 
density and helium abundance. What remains when these dependences
are removed
are $\Gamma_1$ differences caused by differences in the equation of
state and by differences in the heavy-element abundances. For convenience
we call these the $\Gamma_{1, \rm int}$ differences. These differences
do not depend on differences in structure or helium abundance.
We first use a series of models to test the sensitivity of the inversions
and also to determine separately what type of 
$\Gamma_{1, \rm int}$ differences are caused by differences in the
total metallicity, i.e. $Z/X$, differences in the equation of state
and differences in the relative abundances of heavy elements.
From the study of models we find that our inversion results are
reliable in the region $0.77 \le r \le 0.92R_\odot$.  In this
regions $Z/X$ differences cause an almost parallel shift of the
$\delta\Gamma_{1,\rm int}$ curve. EOS differences appear only for 
$r > 0.90 R_\odot$. Enhancing the abundance of one element
gives
humps at different radii, however, if two or more elements are changed
the features can largely cancel out in some cases.
Because of these effects,
it is not possible to distinguish between the relative heavy-element
abundances of the GS98 and the AGS05 mixtures.

Using solar data we find that models constructed with $Z/X=0.0165$,
i.e., the total metallicity of AGS05, are not consistent with the
Sun. The $\Gamma_{1, \rm int}$ for these models are lower than
that of the Sun in the region of the convection zone that we
can study. Models with $Z/X=0.023$, i.e., GS98 total metallicity
fare much better
and confirm the results of Antia \& Basu (2006), who used the
dimensionless sound speed gradient inside the convection zone.
There are however, discrepancies between the $Z/X=0.023$ models
and the Sun close to the surface,
which probably point to differences in EOS between our models and the Sun.
These results thus add to the list of discrepancies between models with
AGS05 metallicities and the seismic data. As mentioned earlier,
the discrepancy between the position of the convection-zone base
was the first to be noted. Since then we know that the models
are also deficient in the core (Basu et al.~2007; Zaatri et al.~2007), and also in
the dimensionless gradient of the sound speed in the convection
zone (Antia \& Basu 2006). Now we can add the discrepancy in
$\Gamma_1$ in the convection zone to that list. It may be noted that
the discrepancy in the depth of the convection zone, the helium abundance
in the convection zone or in the core structure are due to differences in
opacity caused by reduction in $Z$. In principle, these can be resolved
if the opacities are revised upwards by 10--20\% (Basu \& Antia 2004;
Turck-Chi\`eze et al.~2004; Bahcall et al.~2005a).
However, the discrepancies in the dimensionless
sound speed gradient or $\Gamma_1$ will not be affected by increasing
the opacity. Thus it is clear that revision of opacity of solar material
will not resolve all discrepancies. While the dimensionless sound speed gradient
does depend weakly on the density profile or convection zone depth, the
difference $\delta\Gamma_{1,\rm int}$ studied in this work is obtained by
correcting for these
differences and hence depends only on the EOS and heavy-element abundances.
It may be argued that changing $Z$ would also change $X$ or $Y$, which could
indirectly affect $\Gamma_1$. This is not expected since in the region that
we are studying,  $r<0.92R_\odot$, hydrogen and helium are fully ionized
and will not contribute to departure of $\Gamma_1$ from 5/3. The exact
location of ionization zones of various elements depends on treatment of
EOS and there could be some error in these due to uncertainties in EOS.
However, since two independent EOS are giving similar results we expect
that these errors are not too large.
Judging by the nature
of the $\Gamma_{1, \rm int}$ differences between AGS05 models
and the Sun, EOS errors are unlikely to explain the
difference.
From our results we can estimate the $Z/X$ value in the Sun to be around
$0.023$.

Looking at the differences in intrinsic $\Gamma_1$ (Fig.~4) it appears
that the discrepancy in the region $r>0.85R_\odot$ is probably due to
EOS. We may expect the EOS discrepancy to reduce in deeper layers as
we move away from the HeII ionization zone. In view of this discrepancy
it may be difficult to determine the value of $Z/X$, but a value close
to 0.023 appears to be preferred over smaller values. 
It
is possible to further decompose $\delta\Gamma_{1,{\rm int}}$ in terms
of EOS contribution and another contribution from $Z$ or even $Z_i$
the abundances of each heavy element. 
Because of the discrepancy
in EOS and the fact that the ionization zones of different heavy elements
overlap, it may be difficult to determine individual $Z_i$, but it may
still be possible to determine total $Z$.

Since an increased neon abundance has been suggested (Antia \& Basu 2005;
Bahcall et al.~2005b) as a possible
solution to resolve the discrepancy between solar models and seismic
data, we examine the models with enhanced neon abundance.
We find that the models with the abundance of neon enhanced with
respect to the AGS05 values do better than the AGS05 models, but the
improvement appears to be a result of an increase in $Z/X$ rather than
an increase in the neon abundance --- increasing neon
while keeping $Z/X$ fixed makes the agreement between the Sun and models worse.
Thus it appears that increasing neon abundance alone does not help in
resolving the discrepancy. It is necessary to increase the abundances
of other elements like oxygen too. Increasing Ne abundance over the
GS98 abundances may lead to better agreement with seismic data.
But it is difficult to make any definitive statement as it will also
depend on the EOS. It may help if more sophisticated EOS tables are
available for different element mixtures.

Our investigation is handicapped by the lack of high degree modes in the
set. Differences in $\Gamma_{1,\rm int}$ caused by errors in the
equation of state or heavy-element abundances are largest above $0.95R_\odot$.
High-precision frequencies of high degree modes are needed to investigate
the effects of equation of state and abundances on the adiabatic index
properly. However, the study indicates that the heavy-element abundance
of the Sun is significantly higher than that suggested by Asplund et al. (2005) and
close to that of Grevesse \& Sauval (1998).

\section*{Acknowledgments}

We would like to thank the Referee for comments that have led to
improvements in the paper.
We thank the OPAL group for the online opacity tables with different
heavy-element mixtures and the updated EOS tables.
This work utilizes data from the Solar Oscillations
Investigation / Michelson Doppler Imager (SOI/MDI) on the Solar
and Heliospheric Observatory (SOHO).  SOHO is a project of
international cooperation between ESA and NASA.
MDI is supported by NASA grant NAG5-8878
to Stanford University.
This work is partially supported by NSF grant ATM 0348837 to SB.

\clearpage

%---------------------------------------------------------------------
\begin{table}
\caption{OPAL and CEFF Models} 
\label{tab:ceffmdl}
  \begin{center}
{\small
    \begin{tabular}{lccccp{5 true cm}} \\ \hline
      Model Name & $Z/X$ & $Y$ & $r_b/R_\odot$ & Mixture & Enhancements  \\
      \hline
    \multicolumn{6}{l}{{Models with OPAL EOS}} \\
      165OPAL & 0.0165 & 0.2151 & 0.7252 & OpalMix$^1$ &  \\
      245OPAL & 0.0245 & 0.2496 & 0.7153 & OpalMix$^1$ &  \\
    \multicolumn{6}{l}{{Models with CEFF EOS}} \\
      OM165 & 0.0165 & 0.2147 & 0.7284 & OpalMix$^1$ &  \\
      OM245 & 0.0245 & 0.2494 & 0.7181 & OpalMix$^1$ &  \\
      GS245 & 0.0245 & 0.2571 & 0.7203 & GS98$^2$ &  \\
      GS230 & 0.0230 & 0.2514 & 0.7218 & GS98$^2$ &  \\
      GS165 & 0.0165 & 0.2216 & 0.7303 & GS98$^2$ &  \\
      AGS165 & 0.0165 & 0.2313 & 0.7319 & AGS05$^3$ &  \\
      AGS209 & 0.0209 & 0.2530 & 0.7258 & AGS05$^3$ &  \\
      A209Ne2 & 0.0209 & 0.2511 & 0.7220 & AGS05*$^4$ & $2 \times$Ne\\
      A209Ne4 & 0.0209 & 0.24771 & 0.7171 &  AGS05*$^4$ & $4 \times$Ne\\
      A197Ne2CNO & 0.0197 & 0.2394 & 0.7219 & AGS05*$^4$ & $2\times$Ne;
                              enhance C,N,O by 1$\sigma$\\
      GC2 & 0.0245 & 0.2418 & 0.7210 & GS98*$^5$ & $2 \times$C\\
      GO2 & 0.0245 & 0.2291 & 0.7154 & GS98*$^5$ & $2\times$O\\
      GNe2 & 0.0245 & 0.2554 & 0.7156 & GS98*$^5$ & $2 \times$Ne \\
      GO1.5C1.5 & 0.0245 & 0.2344 & 0.7176 & GS98*$^5$ & $1.5\times$(C, O) \\
      \hline
      \end{tabular}
}
  \end{center}
{\small
$^1$ OpalMix: mixture used in OPAL EOS

$^2$ GS98: mixture of \citet{GS1998}

$^3$ AGS05: mixture of \citet{AGS2005}

$^4$ AGS05*: AGS05 with enhancements of certain elements as
noted under column `Enhancements'

$^5$ GS98*: GS98 with enhancements of certain elements as
noted under column `Enhancements'

}
\end{table}
%---------------------------------------------------------------------

\clearpage

\clearpage

%-----------------------------------
%-------------------------

\clearpage

\clearpage

%%%%%%%%%%%%%%%%%% solar data %%%%%%%%%

%%%%%%%%%%%%%%%%%%%%%%%%%%%%%%%%%%%%%%%%%%%%%%%%%%%%%%%%%%%%%%%%%%%%%%%%%%%

\clearpage

\begin{figure}
\centering
\includegraphics[height=0.8\linewidth,angle=90]{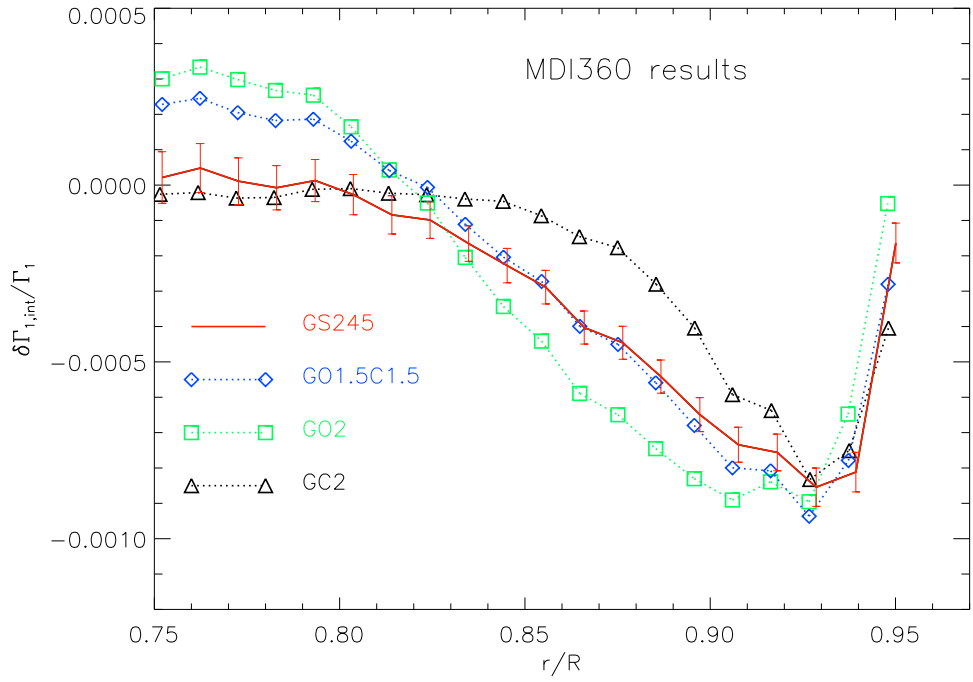}
\caption{
The $\Gamma_{1, \rm int}$ differences between the Sun and models with the abundance of
some elements changed. The value of $Z/X$ is the same for all models.
}
\label{fig:f5}
\end{figure}

\clearpage

\begin{figure}
\centering
\includegraphics[height=0.8\linewidth]{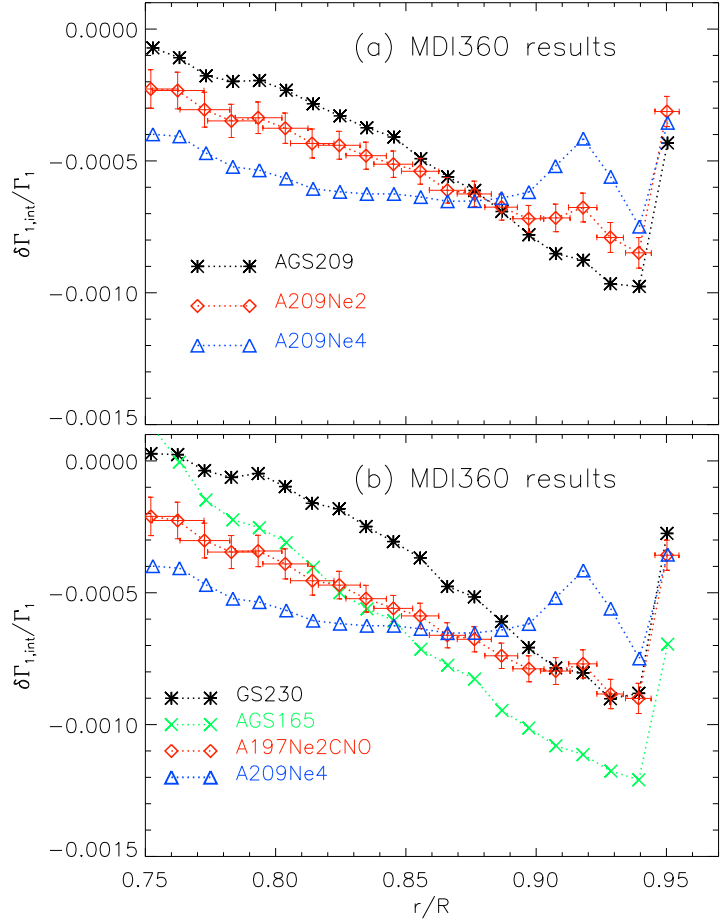}
\caption{The effect of neon enhancement on the $\Gamma_{1, \rm int}$ differences 
between the Sun and models. (a) The effect of enhancing neon abundance when $Z/X$ is
kept the same. (b) The effect of changing $Z/X$ as the neon abundance is enhanced.
The two neon-enhanced models are the ones proposed by Antia \& Basu (2005) to
get models with the AGS05 abundances to agree with the Sun. A normal AGS05 model
and a GS98 model is shown for comparison.
}
\label{fig:f6}
\end{figure}

\end{document}